\documentclass[final]{midl} 

\usepackage{mwe} 
\usepackage{xurl}

\jmlrproceedings{MIDL}{Medical Imaging with Deep Learning}
\jmlrpages{}
\jmlryear{2023}

\title[High-resolution 3D Maps of Left Atrial Displacements]{High-resolution 3D Maps of Left Atrial Displacements using an Unsupervised Image Registration Neural Network\vspace{-3ex}}

\midlauthor{\Name{Christoforos Galazis} \Email{c.galazis20@imperial.ac.uk}\\
\Name{Anil {Anthony Bharath}} \Email{a.bharath@imperial.ac.uk}\\
\Name{Marta Varela} \Email{marta.varela@imperial.ac.uk}\\
\addr Imperial College London, UK \\ \vspace{-12.0mm}
}

\begin{document}

\maketitle
\vspace{-5.0mm}
\begin{abstract}
Functional analysis of the left atrium (LA) plays an increasingly important role in the prognosis and diagnosis of cardiovascular diseases. Echocardiography-based measurements of LA dimensions and strains are useful biomarkers, but they provide an incomplete picture of atrial deformations. High-resolution dynamic magnetic resonance images (Cine MRI) offer the opportunity to examine LA motion and deformation in 3D, at higher spatial resolution and with full LA coverage. However, there are no dedicated tools to automatically characterise LA motion in 3D. Thus, we propose a tool that automatically segments the LA and extracts the displacement fields across the cardiac cycle. The pipeline is able to accurately track the LA wall across the cardiac cycle with an average Hausdorff distance of $2.51 \pm 1.3~mm$ and Dice score of $0.96 \pm 0.02$.
\end{abstract}

\begin{keywords}
Left Atrial, Image Registration Neural Network, Displacement Field Vector.
\end{keywords}

\vspace{-2.0mm}
\section{Introduction}

The analysis of the anatomy and function of the left atrium (LA) is becoming more important for the prognosis and diagnosis of cardiac conditions such as atrial fibrillation (AF) or heart failure (HF) \cite{hoit2017evaluation,peters2021left}. Structural characteristics of the LA are established atrial disease biomarkers \cite{varela2017novel} and analysis of LA deformations has been explored using speckle-tracking echocardiography \cite{smiseth2022imaging}.
These biomarkers are typically obtained for a single LA view and spatial averages across LA wall regions. Spatiotemporal 3D maps of LA deformation are expected to provide more specific signatures of LA pathology, with greater diagnostic and prognostic value, as has been shown for the left ventricle (LV) \cite{duchateau2020machine}. However, there are currently no publicly available MRI datasets or adequate image analysis tools to extract high-resolution displacement field vector (DFV) maps of the whole LA.

In this paper, we use a novel high-resolution Cine MRI protocol designed specifically for the LA. These Cine MRI offer information about the LA at higher spatial resolution than images of any other existing database. However, given that only a small number of subjects have been imaged with this protocol, we develop and utilize methods for limited number of training images.

\textbf{Aim} We propose the following pipeline to automatically obtain high-resolution 3D DFVs of the LA: 1) A few-shot segmentation network (LA-SNet) of the LA across the cardiac cycle to guide the registration; 2) Extraction of the LA segmentation contour and dilation; 3) An automatic subject-by-subject image registration of the LA contour image (LA-DNet).



\section{Methods}
\subsection{Data}
We use 3D LA Cine MRI bSSFP scans acquired using a novel acquisition protocol \cite{varela2020strain}. In summary, they were acquired in a single breath-hold, with resolution of $1.72 \times 1.72 \times 2.00~mm^3$ and 20 phases across the cardiac cycle. Phase 0 corresponds to cardiac end diastole (smallest LA volume). As proof of concept, we analyse images from six subjects: three healthy volunteers and three subjects with suspected cardiovascular disease.


\vspace{-2.0mm}
\subsection{Preprocessing}
The images are cropped to a size of $96 \times 96 \times 36$ voxels, centered at the LA. Additionally, they are translated such that the LA centroid is stationary across the cardiac cycle and their intensity is min-max normalized. We manually segment the LA across the entire cardiac cycle to use as ground truth. From the segmented data, the contour is extracted and dilated using a 2 voxel radius spherical structure, which is used to mask the images.

\vspace{-2.0mm}
\subsection{Model}
\begin{figure}[!ht]
\centering
  \includegraphics[width=0.8\textwidth]{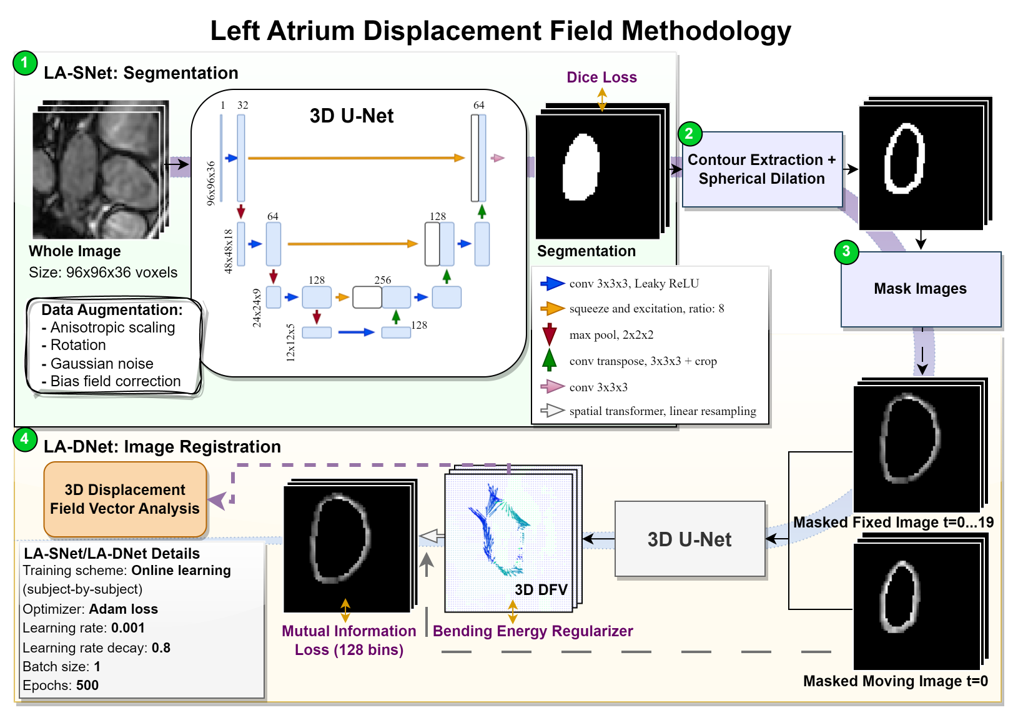}
  \vspace{-5.5mm}
  \caption{The proposed pipeline to extract high-resolution LA displacement field maps.}
  \label{fig:model}
  \vspace{-9.00mm}
\end{figure}

Details of LA-SNet and LA-DNet are in Figure \ref{fig:model}, which their parameters have been experimentally selected. They share the same architecture that is based on a 3D U-Net \cite{ronneberger2015u}. The models incorporate squeeze and excitation blocks \cite{hu2018squeeze}, which were already applied to LV MRI segmentation and image registration \cite{galazis2022tempera}. LA-DNet also utilizes a spatial transformer \cite{jaderberg2015spatial} to obtain the DFV in an unsupervised way. The DFV is smoothed using a bending energy regularizer \cite{rueckert1999nonrigid}. LA-SNet is trained on the augmented whole LA images on cardiac phases 0, 8, and 15 and predicts the respective LA segmentation. LA-DNet takes the two contour masked images (moving: cardiac phase 0; fixed: cardiac phase [0-19]) to generate a displacement field that resamples the moving to the target image.



\vspace{-2.0mm}
\section{Results}
\begin{figure}[!ht]
\centering
  \includegraphics[width=1.0\textwidth]{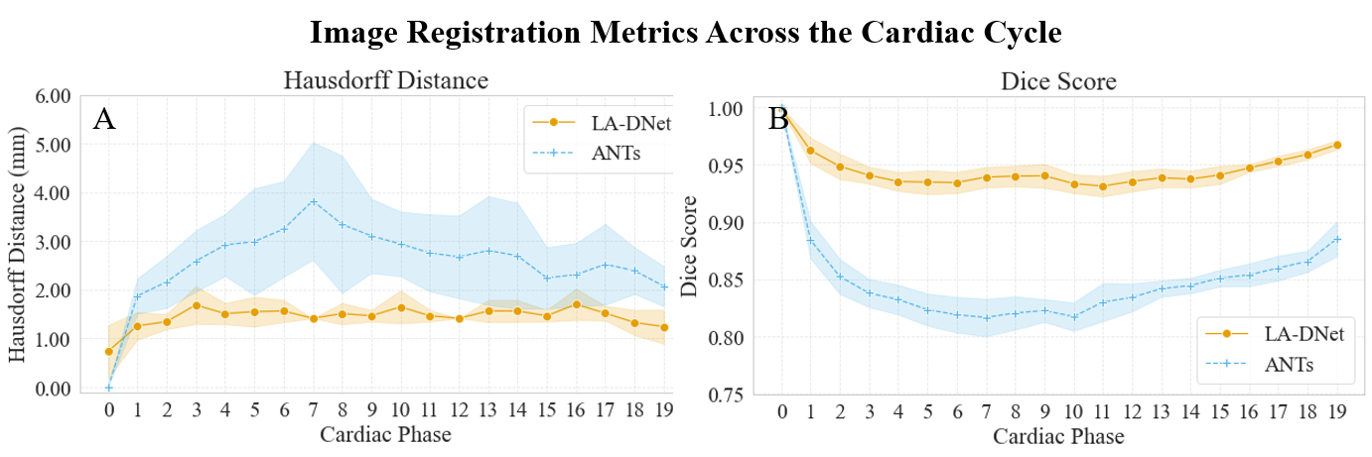}
  \vspace{-8.5mm}
  \caption{The image registration metrics plotted for LA-DNet and ANTs: A) Hausdorff distance (HD) and, B) Dice score (DS). HD and DS are obtained by comparing manual LA segmentations across the cardiac cycle with segmentations transformed using the estimated DFV on phase 0.}
  \label{fig:results}
  \vspace{-9.00mm}
\end{figure}

LA-SNet can accurately segment the LA across the cardiac cycle, with an average Hausdorff distance (HD) of $3.03 \pm 1.12~mm$ and Dice score (DS) of $0.95 \pm 0.02$. Similarly, LA-DNet is able to accurately track the LA wall across the cycle (see Figure \ref{fig:results}). The LA segmentations obtained when adding the estimated DFV to the LA segmentation in phase 0 compare extremely well with the GT segmentations: $HD = 2.51 \pm 1.3~mm; DS = 0.96 \pm 0.02$. It outperformed previously used symmetric diffeomorphic image normalization from ANTs package \cite{avants2009advanced} which obtained $(HD = 2.57 \pm 1.16~mm; DS =0.85 \pm 0.04)$ to the same LA contour images and $(HD = 3.35 \pm 1.48~mm; DS =0.77 \pm 0.09)$ when applied to the unsegmented LA images. Using LA-DNet directly on the unsegmented LA images as inputs also led to poor results $(HD = 3.35 \pm 1.05~mm; DS =0.78 \pm 0.07)$. The LA-DNet estimated DFVs are spatially and temporally smoother and the Jacobian of the deformation gradient is consistent with the known volumetric changes of the LA, as can be seen in: \href{https://datalore.jetbrains.com/view/notebook/IEjxZeluEDQ6bkbI8SfqJv}{https://tinyurl.com/2eju3r9f}.

\vspace{-4.00mm}

\vspace{-3.0mm}
\section{Conclusions}
The proposed pipeline is able to extract DFVs that accurately track the LA wall across the cardiac cycle. The estimated high-resolution 3D LA DFVs pave the way towards potentially detecting regional functional biomarkers for conditions such as AF or HF. They may also provide useful information for the identification of LA fibrosis \cite{sohns2020atrial}.

The LA registration across the cardiac cycle is more challenging than that of the LV. For the latter, several registration tools are available \cite{hernandez2021deep,de2019deep}, but these performed poorly for the LA registration task. The usual assumption, that the intensity of the different image components (e.g. the LV myocardium) is constant across the cardiac cycle, is not valid for the LA. This is because the LA myocardium is very thin \cite{varela2017novel}, and thus barely identifiable in bSSFP images; and the LA blood pool voxels' intensity depends on blood velocity and is therefore very variable across the cardiac cycle. We successfully propose a different approach for automatically LA registration, using LA contours from automated segmentations as inputs, training it on a subject by subject basis to allow its deployment to small datasets of Cine MRI of the LA.

\midlacknowledgments{This work was supported by the UKRI CDT in AI for Healthcare \url{http://ai4health.io} (Grant No. EP/S023283/1) and the British Heart Foundation Centre of Research Excellence at Imperial College London (RE/18/4/34215). We acknowledge computational resources and support provided by the Imperial College Research Computing Service (\url{http://doi.org/10.14469/hpc/2232}). Last but not least, we thank the volunteers for allowing the use of their data for this research.}

\bibliography{midl-shortpaper}

\begin{thebibliography}{15}
\providecommand{\natexlab}[1]{#1}
\providecommand{\url}[1]{\texttt{#1}}
\expandafter\ifx\csname urlstyle\endcsname\relax
  \providecommand{\doi}[1]{doi: #1}\else
  \providecommand{\doi}{doi: \begingroup \urlstyle{rm}\Url}\fi

\bibitem[Avants et~al.(2009)Avants, Tustison, Song, et~al.]{avants2009advanced}
Brian~B Avants, Nick Tustison, Gang Song, et~al.
\newblock Advanced normalization tools (ants).
\newblock \emph{Insight j}, 2\penalty0 (365):\penalty0 1--35, 2009.

\bibitem[De~Vos et~al.(2019)De~Vos, Berendsen, Viergever, Sokooti, Staring, and
  I{\v{s}}gum]{de2019deep}
Bob~D De~Vos, Floris~F Berendsen, Max~A Viergever, Hessam Sokooti, Marius
  Staring, and Ivana I{\v{s}}gum.
\newblock A deep learning framework for unsupervised affine and deformable
  image registration.
\newblock \emph{Medical image analysis}, 52:\penalty0 128--143, 2019.

\bibitem[Duchateau et~al.(2020)Duchateau, King, and
  De~Craene]{duchateau2020machine}
Nicolas Duchateau, Andrew~P King, and Mathieu De~Craene.
\newblock Machine learning approaches for myocardial motion and deformation
  analysis.
\newblock \emph{Frontiers in cardiovascular medicine}, 6:\penalty0 190, 2020.

\bibitem[Galazis et~al.(2022)Galazis, Wu, Li, Petri, Bharath, and
  Varela]{galazis2022tempera}
Christoforos Galazis, Huiyi Wu, Zhuoyu Li, Camille Petri, Anil~A Bharath, and
  Marta Varela.
\newblock Tempera: Spatial transformer feature pyramid network for cardiac mri
  segmentation.
\newblock In \emph{Statistical Atlases and Computational Models of the Heart.
  Multi-Disease, Multi-View, and Multi-Center Right Ventricular Segmentation in
  Cardiac MRI Challenge: 12th International Workshop, STACOM 2021, Held in
  Conjunction with MICCAI 2021, Strasbourg, France, September 27, 2021, Revised
  Selected Papers}, pages 268--276. Springer, 2022.

\bibitem[Hernandez et~al.(2021)Hernandez, Rienm{\"u}ller, Baumgartner, and
  Baumgartner]{hernandez2021deep}
Karen Andrea~Lara Hernandez, Theresa Rienm{\"u}ller, Daniela Baumgartner, and
  Christian Baumgartner.
\newblock Deep learning in spatiotemporal cardiac imaging: A review of
  methodologies and clinical usability.
\newblock \emph{Computers in Biology and Medicine}, 130:\penalty0 104200, 2021.

\bibitem[Hoit(2017)]{hoit2017evaluation}
Brian~D Hoit.
\newblock Evaluation of left atrial function: current status.
\newblock \emph{Structural Heart}, 1\penalty0 (3-4):\penalty0 109--120, 2017.

\bibitem[Hu et~al.(2018)Hu, Shen, and Sun]{hu2018squeeze}
Jie Hu, Li~Shen, and Gang Sun.
\newblock Squeeze-and-excitation networks.
\newblock In \emph{Proceedings of the IEEE conference on computer vision and
  pattern recognition}, pages 7132--7141, 2018.

\bibitem[Jaderberg et~al.(2015)Jaderberg, Simonyan, Zisserman,
  et~al.]{jaderberg2015spatial}
Max Jaderberg, Karen Simonyan, Andrew Zisserman, et~al.
\newblock Spatial transformer networks.
\newblock \emph{Advances in neural information processing systems}, 28, 2015.

\bibitem[Peters et~al.(2021)Peters, Lamy, Sinusas, and
  Baldassarre]{peters2021left}
Dana~C Peters, J{\'e}r{\^o}me Lamy, Albert~J Sinusas, and Lauren~A Baldassarre.
\newblock Left atrial evaluation by cardiovascular magnetic resonance:
  sensitive and unique biomarkers.
\newblock \emph{European Heart Journal-Cardiovascular Imaging}, 23\penalty0
  (1):\penalty0 14--30, 2021.

\bibitem[Ronneberger et~al.(2015)Ronneberger, Fischer, and
  Brox]{ronneberger2015u}
Olaf Ronneberger, Philipp Fischer, and Thomas Brox.
\newblock U-net: Convolutional networks for biomedical image segmentation.
\newblock In \emph{Medical Image Computing and Computer-Assisted
  Intervention--MICCAI 2015: 18th International Conference, Munich, Germany,
  October 5-9, 2015, Proceedings, Part III 18}, pages 234--241. Springer, 2015.

\bibitem[Rueckert et~al.(1999)Rueckert, Sonoda, Hayes, Hill, Leach, and
  Hawkes]{rueckert1999nonrigid}
Daniel Rueckert, Luke~I Sonoda, Carmel Hayes, Derek~LG Hill, Martin~O Leach,
  and David~J Hawkes.
\newblock Nonrigid registration using free-form deformations: application to
  breast mr images.
\newblock \emph{IEEE transactions on medical imaging}, 18\penalty0
  (8):\penalty0 712--721, 1999.

\bibitem[Smiseth et~al.(2022)Smiseth, Baron, Marino, Marwick, and
  Flachskampf]{smiseth2022imaging}
Otto~A Smiseth, Tomasz Baron, Paolo~N Marino, Thomas~H Marwick, and Frank~A
  Flachskampf.
\newblock Imaging of the left atrium: pathophysiology insights and clinical
  utility.
\newblock \emph{European Heart Journal-Cardiovascular Imaging}, 23\penalty0
  (1):\penalty0 2--13, 2022.

\bibitem[Sohns and Marrouche(2020)]{sohns2020atrial}
Christian Sohns and Nassir~F Marrouche.
\newblock Atrial fibrillation and cardiac fibrosis.
\newblock \emph{European heart journal}, 41\penalty0 (10):\penalty0 1123--1131,
  2020.

\bibitem[Varela et~al.(2017)Varela, Bisbal, Zacur, Berruezo, Aslanidi, Mont,
  and Lamata]{varela2017novel}
Marta Varela, Felipe Bisbal, Ernesto Zacur, Antonio Berruezo, Oleg~V Aslanidi,
  Lluis Mont, and Pablo Lamata.
\newblock Novel computational analysis of left atrial anatomy improves
  prediction of atrial fibrillation recurrence after ablation.
\newblock \emph{Frontiers in physiology}, 8:\penalty0 68, 2017.

\bibitem[Varela et~al.(2020)Varela, Queir{\'o}s, Anjari, Correia, King,
  Bharath, and Lee]{varela2020strain}
Marta Varela, Sandro Queir{\'o}s, Mustafa Anjari, Teresa Correia, Andrew~P
  King, Anil~A Bharath, and Jack Lee.
\newblock Strain maps of the left atrium imaged with a novel high-resolution
  cine mri protocol.
\newblock In \emph{2020 42nd Annual International Conference of the IEEE
  Engineering in Medicine \& Biology Society (EMBC)}, pages 1178--1181. IEEE,
  2020.

\end{thebibliography}

\end{document}